\documentclass[11pt,a4paper]{article}
\usepackage[hyperref]{acl2019}
\usepackage{times}
\usepackage{latexsym}
\usepackage{graphicx}
\usepackage[frozencache]{minted}
\AtBeginEnvironment{minted}{%
  }

\usepackage{url}

\aclfinalcopy 
\def\aclpaperid{66} 

\title{GrapAL: Connecting the Dots in Scientific Literature}

\author{Christine Betts, Joanna Power, Waleed Ammar \\
  Allen Institute for Artificial Intelligence, Seattle, WA, USA \\
  {\tt chrstn@cs.washington.edu,  \{joannap,waleeda\}@allenai.org} \\}

\date{}

\begin{document}
\definecolor{green}{HTML}{005B83}
\definecolor{purple}{HTML}{4C3FAA}

\maketitle
\begin{abstract}
    We introduce GrapAL (Graph database of Academic Literature), a versatile tool for exploring and investigating a knowledge base of scientific literature, that was semi-automatically constructed using NLP methods. GrapAL satisfies a variety of use cases and information needs requested by researchers.
    At the core of GrapAL is a Neo4j graph database with an intuitive schema and a simple query language.
    In this paper, we describe the basic elements of GrapAL, how to use it, and several use cases such as finding experts on a given topic for peer reviewing, discovering indirect connections between biomedical entities and computing citation-based metrics. 
    We open source the demo code to help other researchers develop applications that build on GrapAL.\footnote{\url{https://github.com/allenai/grapal-website}}
\end{abstract}

\section{Introduction}

Researchers rely on scientific literature to perform a wide variety of tasks such as searching for papers, assessing applicants for a research position and keeping track of papers published on topics of interest.
Several software tools are available to help researchers perform these tasks.
For example, 
many biomedical researchers use PubMed to find papers relevant for their studies,\footnote{\url{https://www.ncbi.nlm.nih.gov/pubmed/}}
Google Scholar allows researchers to verify and curate their user profiles,\footnote{\url{https://scholar.google.com/}}
and Semantic Scholar extracts research topics, figures, and tables from papers and links them to external content such as slides, videos and GitHub repositories.\footnote{\url{https://www.semanticscholar.org/}}
However, such tools tend to only feature the most commonly used functionalities in order to keep the interface simple for users, ignoring the long tail of informational needs such as finding experts on a given topic, identifying potential collaborators, assessing influence between research areas, and discovering connections between biological entities.

In this paper, we address these limitations by introducing a tool that provides a flexible and efficient way to query the Semantic Scholar knowledge base, an automatically constructed knowledge base of scientific literature \cite{Ammar2018ConstructionOT}.
In addition to bridging the gap between available tools and informational needs of researchers, GrapAL demonstrates how automatically constructed knowledge bases can be effectively used to solve problems in the real world.

GrapAL is publicly available at \url{grapal.allenai.org}, along with documentation.\footnote{A screencast of the tool is available at \url{https://www.youtube.com/watch?v=1ivX9sHw2RU&feature=youtu.be}}
In the following section (\S\ref{sec:howto}), we introduce the schema and query language used in GrapAL and discuss how users can connect to the database.
In \S\ref{sec:case_studies}, we show how GrapAL can be used to satisfy several example informational needs identified through user studies.
In \S\ref{sec:system}, we discuss some of the design choices and the system architecture for GrapAL.

\section{How to Use GrapAL}\label{sec:howto}
GrapAL is designed to satisfy a wide variety of use cases and scenarios requested by users of Semantic Scholar who need to process scientific literature in order to facilitate their own work.
To achieve this, we design GrapAL as a Neo4j property graph with an intuitive schema, and make it available to query using the Cypher query language \cite{Francis2018CypherAE}.

\begin{figure*}
\includegraphics[width=1.0\textwidth]{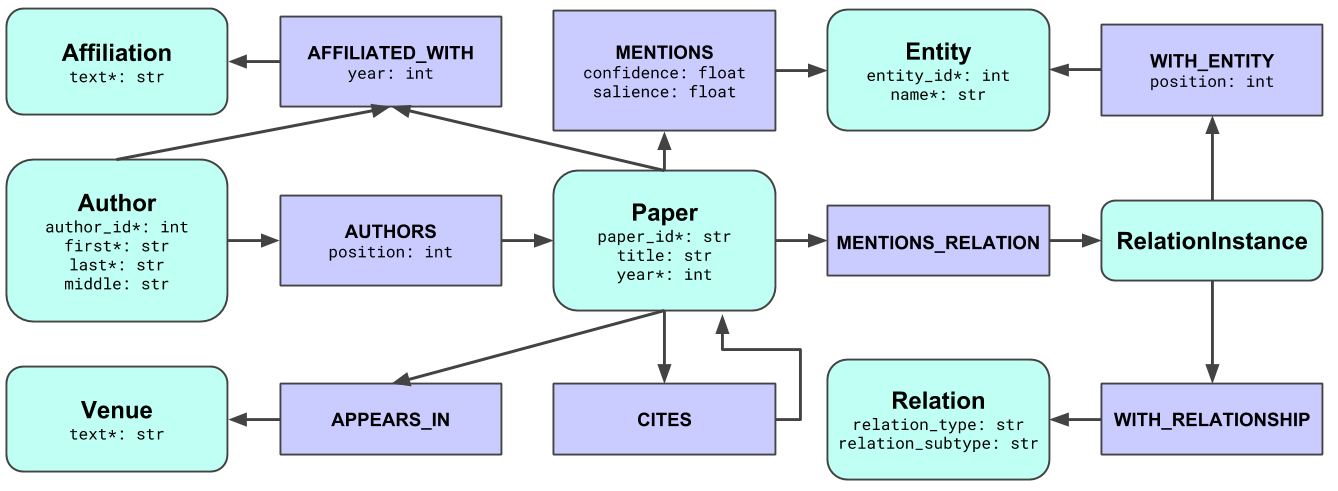}
\caption{Overview of GrapAL schema. *denotes indexed property.  }\label{fig:schema}
\end{figure*}

\paragraph{Schema.}
Fig.~\ref{fig:schema} demonstrates the schema of our graph database, which consists of 7 node types (displayed in turquoise) and 8 edge types (displayed in purple).
The properties associated with each node and edge type are listed.
In order to avoid violating intellectual property of publishers, we do not include some information about papers such as the abstract and full text.

At the core of the graph is the \texttt{\color{green}Paper} node.
\texttt{\color{green}Paper} nodes may connect to \texttt{\color{green}Venue} nodes, \texttt{\color{green}Author} nodes, 
\texttt{\color{green}Affiliation} nodes, 
\texttt{\color{green}Entity} nodes,
\texttt{\color{green}RelationInstance} nodes
or other \texttt{\color{green}Paper} nodes via \texttt{\color{purple}APPEARS\_IN} edges, 
\texttt{\color{purple}AUTHORS} edges,
\texttt{\color{purple}AFFILIATED\_WITH} edges,
\texttt{\color{purple}MENTIONS} edges,
\texttt{\color{purple}MENTIONS\_RELATION} edges and \texttt{\color{purple}CITES} edges, respectively.
A \texttt{\color{green}RelationInstance} node, e.g., \textsc{Causes[Smoking,Cancer]}, represents an n-ary relationship of type \texttt{\color{green}Relation} (via a \texttt{\color{purple}WITH\_RELATIONSHIP} edge) between two or more \texttt{\color{green}Entity} nodes (via \texttt{\color{purple}WITH\_ENTITY} edges).
Details on how we extract entities and various metadata for each paper can be found in \citet{Ammar2018ConstructionOT}.
The only schema changes introduced in this work are including \texttt{\color{green}Affiliation} and \texttt{\color{green}Venue} nodes (and corresponding edge types), and optimizing for query execution time.
Table \ref{tab:counts} provides the number of instances of each node and edge type in the schema at the time of this writing.

\begin{table}[t!]
\scriptsize
\begin{center}
    \begin{minipage}{.5\linewidth}
      \centering
\begin{tabular}{lr}
 \bf Node Type & \bf Count \\ \hline
\texttt{Affiliation}* & 16M \\
\texttt{Author} & 17M \\
\texttt{Entity} & 493K \\
\texttt{Paper} & 46M  \\
\texttt{Relation} & 51 \\
\texttt{RelationInstance} & 347K\\
\texttt{Venue}* & 78K \\
\\
\end{tabular}
    \end{minipage}%
    \begin{minipage}{.5\linewidth}
      \centering
      \begin{tabular}{|lr}
 \bf Edge Type & \bf Count \\ \hline
\texttt{AFFILIATED\_WITH} & 119M \\
\texttt{APPEARS\_IN} & 67M \\
\texttt{AUTHORS} & 148M \\
\texttt{CITES} & 693M \\
\texttt{MENTIONS} & 400M \\
\texttt{MENTIONS\_RELATION} & 73M\\
\texttt{WITH\_ENTITY} & 1M \\
\texttt{WITH\_RELATIONSHIP} & 350K\\
\end{tabular}
    \end{minipage} 
\end{center}
\caption{\label{tab:counts} Approximate node and edge cardinalities. (*) indicates node types that are not canonicalized.}
\end{table}

\paragraph{Query Language.}
Before we discuss realistic case studies in \S\ref{sec:case_studies}, we introduce the query language used in GrapAL with a few toy examples:

First, consider the following query that matches arbitrary author nodes in GrapAL and returns the first 10:

\begin{minted}[tabsize=2,breaklines,fontsize=\small]{cypher}
// Find arbitrary authors.
MATCH (a:Author) RETURN a LIMIT 10
\end{minted}

More often than not, we only want to match nodes with some desired properties. In the next example, we only match authors with first name `Clarence' and last name `Ellis'. Note the round brackets used to specify an instance of node type \texttt{Author}, and the curly brackets used to specify its properties.

\begin{minted}[tabsize=2,breaklines,fontsize=\small]{cypher}
// Find authors by name.
MATCH (a:Author {last: "Ellis", first: "Clarence"})
RETURN a
\end{minted}

Alternatively, we could use a \texttt{WHERE} clause to specify the desired properties of matched nodes, as demonstrated in the following example that matches papers by their title.
This example also shows how to match nodes by specifying their relation to another node, e.g., authors of a paper.
Note the use of square brackets to specify edges and the arrow to specify edge direction.

\begin{minted}[tabsize=2,breaklines,fontsize=\small]{cypher}
// Find authors of a specific paper.
MATCH (a:Author)-[:AUTHORS]->(p:Paper)
WHERE p.title = "One-shot learning of object categories"
RETURN a
\end{minted}

More information about the Cypher query language can be found in \citet{Francis2018CypherAE}.

\paragraph{Connecting to GrapAL.}
Users can query GrapAL in a variety of methods.
First, an interactive graphical interface is available at \url{https://grapal.allenai.org:7473/browser/} that is suitable for interactive exploration of GrapAL with a relatively small number of results.
We demonstrate how the interactive interface could be used in a screencast.\footnote{\url{https://www.youtube.com/watch?v=1ivX9sHw2RU&feature=youtu.be}}

Users can also build web applications that leverage GrapAL by making RESTful queries to an HTTP API.\footnote{Documentation for the API is available at \url{https://neo4j.com/docs/rest-docs/current/}}
As an example, we have developed a simple web-based application at \url{https://grapal.allenai.org/app} that can be used to load any of the case studies described in the next section.\footnote{ For example, the following URL will load the shortest path example: \url{https://grapal.allenai.org/app/?example=shortest-path}}
Users can also type in arbitrary queries, share the queries with collaborators, and download the results in JSON format.

Users can also query the graph natively in their favourite programming language using one of the Neo4j language drivers.
Neo4j officially supports five languages: .NET, Java, Javascript, Go and Python, but drivers are available for a longer list of programming languages thanks to the active Neo4j community.\footnote{See \url{https://neo4j.com/developer/language-guides/} for the complete list of Neo4j language drivers.}
We provide an example of using the Python driver to compute disruption scores as described in \citet{Wu2019LargeTD}.\footnote{\url{https://gist.github.com/chrstnb/088f7699930ad53e757906f4d3d6c1f5}}
\paragraph{DOI and ArXivId Compatibility.}
Users can switch between Digital Object Identifiers (DOIs) or arXiv identifiers (ArXivId) and paper IDs with the Semantic Scholar API\footnote{\url{http://api.semanticscholar.org/}}.
For example, we can look up the paper node corresponding to the DOI 10.1038/nrn3241 by first executing the HTTP query \url{https://api.semanticscholar.org/v1/paper/10.1038/nrn3241} that returns a JSON object with paper ID \texttt{931d6b6ee097eab80b8f89a313c8d3a6d 5443cb2}. Then, we execute the Cypher query: 

\begin{minted}[tabsize=2,breaklines,fontsize=\small]{cypher}
// Look up paper by ID.
MATCH (p:Paper {paper_id: "931d6b6ee097eab80b8f89a313c8d3 a6d5443cb2"}) 
RETURN p
\end{minted}
In the future, we plan to add DOI properties and ArXivId properties to the knowledge base as well.

\section{Case Studies}
\label{sec:case_studies}
We interviewed computer science and biomedical researchers to better understand the kinds of questions they would like to answer via a knowledge base of scientific literature. 
In this section, we focus on some of the more compelling use cases that were identified in the interviews, and provide example queries to address them in GrapAL.

For each example we give a link to load the query in the query loader and the full text of the query. From the query loader, users can view or save the results of a query and also copy it to be pasted into the Neo4j browser, where users can view interactive visualizations of the query results.

\begin{figure}
\includegraphics[width=\columnwidth]{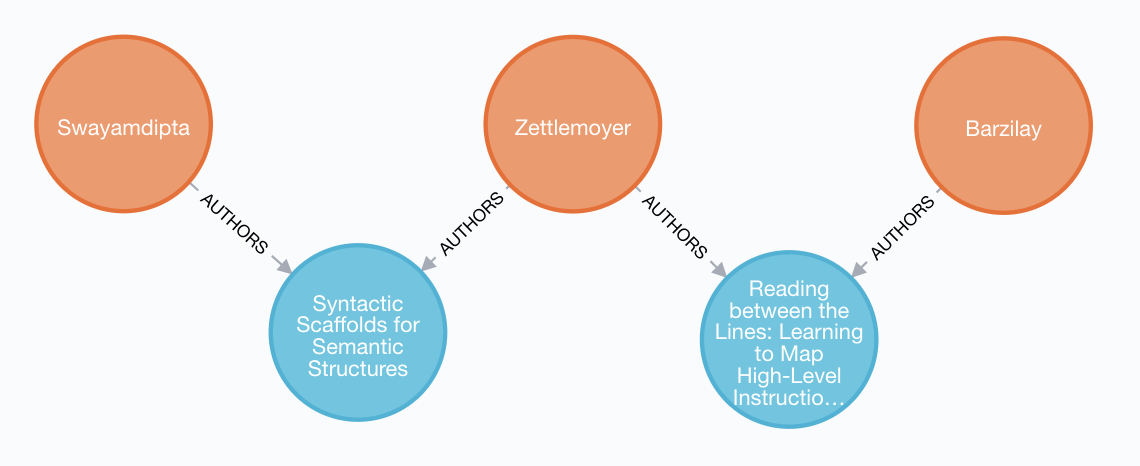}
\caption{Shortest path between Swabha Swayamdipta and Regina Barzilay.\label{fig:swabs}}
\end{figure}

\paragraph{Shortest Path.}
Consider a researcher \textbf{\texttt{a}} seeking an introduction or an endorsement to work with another researcher \textbf{\texttt{b}}.
By finding the shortest path between the two researchers in GrapAL, researcher \textbf{\texttt{a}} can identify common collaborators connecting the two.
The following query, for instance, matches a path connecting Swabha Swayamdipta and Regina Barzilay using authorship edges only, and returns a path that connects them via Luke Zettlemoyer who co-authored papers with both researchers (see Fig.~\ref{fig:swabs}).\footnote{This query can be loaded and modified at \url{https://grapal.allenai.org/app/?example=shortest-path}}
\begin{minted}[tabsize=2,breaklines,fontsize=\small]{cypher}
// Find shortest path between two researchers by name.
MATCH p=shortestPath((a:Author)-
    [:AUTHORS*0..6]-(b:Author)) 
WHERE a.first = "Swabha" 
    AND a.last = "Swayamdipta"
    AND b.first = "Regina" 
    AND b.last = "Barzilay"
RETURN p
\end{minted}

In this example, we constrain the number and type of edges in the graph to a maximum of six \texttt{AUTHORS} edges.
For authors with an ambiguous name, it may be necessary to specify the author by their ID, which can be found by inspecting their author page URL on Semantic Scholar:\footnote{E.g., Swabha Swayamdipta author page URL is \url{https://www.semanticscholar.org/author/Swabha-Swayamdipta/2705113}} 
\begin{minted}[tabsize=2,breaklines,fontsize=\small]{cypher}
// Find shortest path between two researchers, one by author ID.
MATCH p=shortestPath((a:Author)-
    [:AUTHORS*0..6]-(b:Author)) 
WHERE a.author_id = 2705113 
    AND b.first = "Regina" 
    AND b.last = "Barzilay"
RETURN p
\end{minted}
Similar queries can be used to find colleagues who published at a given venue, or currently work at a given university or research lab.

\paragraph{Finding Experts.}
One of the pain points in organizing a conference is identifying reviewers who are knowledgeable about the research topics discussed in submitted papers.
By querying GrapAL, members of the organizing committee will be able to find more competent reviewers, while relying less on their (often biased) professional network when deciding whom to invite for peer reviewing.
For example, the following query can be used to find researchers who published the most on ``Relationship extraction'' since 2013.\footnote{This query can be loaded and modified at \url{https://grapal.allenai.org/app/?example=experts}}
\begin{minted}[tabsize=2,breaklines,fontsize=\small]{cypher}
// Find authors who published the most on relation extraction since 2013.
MATCH (a:Author)-[:AUTHORS]->(p:Paper),
    (p)-[:MENTIONS]->
    (:Entity {name: "Relationship 
        extraction"})
WHERE p.year > 2013
WITH a, count(p) as cp
RETURN a, cp 
ORDER BY cp DESC 
\end{minted}
Here, we use \texttt{ORDER BY cp DESC} to sort the authors by the number of papers they published on this topic.
In order to find the node that represents a topic of interest in GrapAL, users could use the search feature on semantic scholar and inspect the relevant topic page URL for the entity ID, or use regular expressions to query GrapAL, e.g.,\footnote{This query can be loaded and modified at \url{https://grapal.allenai.org/app/?example=canonical-entity}}
\begin{minted}[tabsize=2,breaklines,fontsize=\small]{cypher}
// Fuzzy matching of entity names.
MATCH (e:Entity)
WHERE e.name =~ "(?i)relationship 
    extraction"
RETURN e 
\end{minted}

\begin{figure}
\includegraphics[width=\columnwidth]{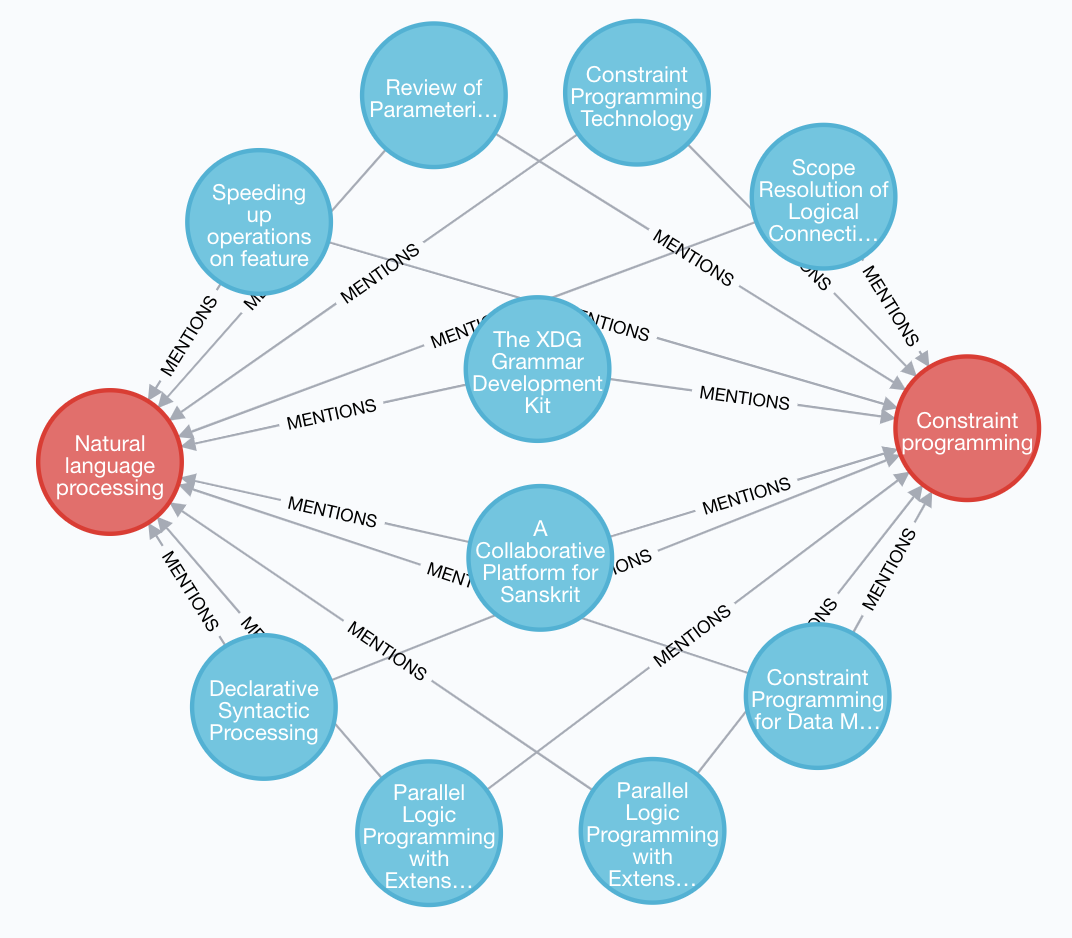}
\caption{Ten papers that mention both `Natural language processing' and `Constraint programming.'.\label{fig:constraint}}
\end{figure}

\paragraph{Papers at the Intersection of Entities.}
Search engine results sometimes make it difficult to find papers that discuss multiple topics or fields. With GrapAL, we can return papers that discuss any number of entities of interest, e.g., "Constraint programming" and "Natural language processing". Fig.~\ref{fig:constraint} shows a visualization of the results on the Neo4j browser, limited to 10 papers.\footnote{This query can be loaded and modified at \url{https://grapal.allenai.org/app/?example=intersecting-entities}}
\begin{minted}[tabsize=2,breaklines,fontsize=\small]{cypher}
// Find papers that mention both constraint programming and natural language processing.
MATCH (p:Paper)-[:MENTIONS]->
    (e1:Entity {name: "Constraint 
        programming"}),
    (p:Paper)-[:MENTIONS]->
    (e2:Entity {name: "Natural language 
        processing"}) 
RETURN p
\end{minted}

\paragraph{Connecting Scientific Concepts.}
Some researchers wanted to explore direct and indirect connections between two scientific concepts (entities) of interest, e.g., the impact of `adjuvant antiestrogen therapy (Arimidex)' on `estrogen receptors'.
Using GrapAL, we can find how two entities are indirectly connected via coded relationships and a chain of entities in the knowledge base, which can help generate new hypotheses or quickly assess the viability of a hypothesis before conducting expensive lab experiments.\footnote{This query can be loaded and modified at \url{https://grapal.allenai.org/app/?example=scientific-concepts}}
\begin{minted}[tabsize=2,breaklines,fontsize=\small]{cypher}
// Find path between Estrogen Receptors and Arimidex via coded relationships.
MATCH path=shortestPath(
    (er:Entity {name: "Estrogen 
        Receptors"})-
    [:WITH_ENTITY*0..15]-
    (ar:Entity {name: "Arimidex"}))
WITH nodes(path) as ns
UNWIND ns as n
MATCH (n)-[:WITH_ENTITY {position: 0}]->
    (e0:Entity),
    (n)-[:WITH_ENTITY {position: 1}]->
        (e1:Entity),
    (n)-[:WITH_RELATIONSHIP]->
        (r:Relation) 
RETURN e0, r, e1
\end{minted}
This query returns a list of triples (e0, r, e1) that connect `Arimidex' to `Estrogen Receptors'.
The \texttt{UNWIND} operator allows us to examine each node on the shortest path and process it as needed. 

\paragraph{Citation-Based Metrics.}
Citations are often used as a proxy for the impact of papers, researchers or venues.
In addition to computing traditional metrics such as h-index and i10-index, GrapAL can also be used to compute more granular metrics, e.g., to estimate the rate at which papers in one conference cite papers in another conference:\footnote{ This query can be loaded and modified at \url{https://grapal.allenai.org/app/?example=citation-metrics}}
\begin{minted}[tabsize=2,breaklines,fontsize=\small]{cypher}
// Find the number of times a NAACL paper cites a CVPR paper.
MATCH (p1:Paper)-[:APPEARS_IN]->
    (naacl:Venue),
(p2:Paper)-[:APPEARS_IN]->(cvpr:Venue),
path=((p1)-[:CITES]->(p2)) 
WHERE naacl.text =~ ".*NAACL.*" 
AND cvpr.text =~ ".*CVPR.*" 
RETURN count(path)
\end{minted}
This query returns the number of times a NAACL paper cites a CVPR paper. 
We use the \texttt{=\~} operator to match on venue names by regular expression because venues are stored as unstructured strings.

\section{System Design}
\label{sec:system}
\paragraph{Graph Database.}
Due to the high connectivity in the data and the nature of queries GrapAL is designed for, we opted to create GrapAL using a graph-native database instead of a more conventional relational database.
Unlike a relational database, a graph database provides a natural and efficient way to query and traverse multi-hop relations without using computationally expensive join operations.
Several graph database systems have recently become available, including AWS Neptune, Grakn.ai, dgraph and Neo4j. 
We decided to build GrapAL on Neo4j since it is one of the more mature platforms, has a strong community of developers, and is the most widely used graph database system as of the time of this writing.\footnote{\url{https://db-engines.com/en/ranking/graph+dbms}}
One limitation of Neo4j is that it is not a distributed database system, but we were able to fit GrapAL on a single server.

\paragraph{Building and Deploying GrapAL.}
GrapAL is powered by the same data that powers the \url{semanticscholar.org} website, as described in \citet{Ammar2018ConstructionOT}.
We use a \textit{staging server} to read a snapshot of the data as Spark DataFrames from AWS S3 and write CSV files that match the property schema described earlier.
Due to the sheer amount of records, we process different shards of the data in parallel before aggregating all shards into one CSV file for each node and edge type of the schema.
Then, we use the Neo4j CSV import function to build the database. Once we've built the database, we start up a Neo4j server and run a Cypher script to create indexes.
The staging server is an EC2 machine with instance type \texttt{r5.24xlarge}. This process takes around 6 hours and the resulting database is roughly 80 GB (including indexes).

Once the data is imported, the database files are copied over to a \textit{production server} that serves the dataset publicly and has lower processor and memory requirements compared to the staging server.
The staging server is an EC2 machine with instance type \texttt{r4.16xlarge}.
We plan to rebuild GrapAL at a monthly cadence with new snapshots of the data.

\section{Related Work}

Related APIs are available to help researchers navigate scientific literature. \citet{Singh2018CLST} provides an API to interact with the ACL anthology. However, it is limited to the areas of computational linguistics and natural langauge processing, and it uses a predefined list of query templates with placeholders for authors, papers and venues. 
Springer Nature SciGraph \footnote{\url{https://scigraph.springernature.com/explorer}} provides an API for accessing publication metadata from the Springer Nature corpus, but it is limited to papers and books published by Springer Nature. 
The Microsoft Academic Graph \cite{Shen2018AWS} is similarly an API for examining academic literature. As a relational database, it is hard to query with complex, multi-hop relations as discussed in \S\ref{sec:system}.
This work is also related to a line of NLP work focusing on scientific documents including citation prediction \cite[e.g.,][]{D11-1055,Bhagavatula2018ContentBasedCR}, author modeling \cite[e.g.,][]{D15-1175}, stylometry \cite[e.g.,][]{N12-1033}, bibliometrics \cite[e.g.,][]{D13-1012,Weihs2017LearningTP} and information extraction \cite[e.g.,][]{L18-1299,L18-1587}.

\section{Conclusion}
GrapAL is a versatile tool for exploring and investigating scientific literature built on the Neo4j graph database framework. We describe the basic elements of GrapAL, how to use it, and use cases such as finding experts on a given topic for peer reviewing, discovering indirect connections between biomedical entities, and computing citation-based metrics.

Future improvements include more metadata and changes to the structure of affiliation and venue data. 
We intend to change the data pipeline architecture to perform event-based incremental updates rather than a regular batch build.
We continue to improve the models used to populate GrapAL's nodes and edges (e.g., author disambiguation and entity extraction and linking).

\section*{Acknowledgments}
We thank Khaled Ammar for his graph database suggestions, Michal Guerquin for his help in designing and building the pipeline, and Darrell Plessas for his technical assistance. We also thank Noah Smith and the Semantic Scholar team for their support. \\

\bibliography{main}
\bibliographystyle{acl_natbib}

\end{document}